# Chirality locking charge density waves in a chiral crystal

Geng Li[1,2,3,4#], Haitao Yang[1,2#], Peijie Jiang[1,2#], Cong Wang[5#], Qiuzhen Cheng[1,2], Shangjie Tian[5], Guangyuan Han[1,2], Chengmin Shen[1,2], Xiao Lin[1,2], Hechang Lei[5*], Wei Ji[5*], Ziqiang Wang[6*], Hong-Jun Gao[1,2,3,4*]

[1] Institute of Physics, Chinese Academy of Sciences, Beijing 100190, China

[2] School of Physical Sciences, University of Chinese Academy of Sciences, Beijing 100190, China

[3] CAS Center for Excellent in Topological Quantum Computation, University of Chinese Academy of Sciences, Beijing 100190, China

[4] Songshan Lake Materials Laboratory, Dongguan, Guangdong 523808, PR China

[5] Beijing Key Laboratory of Optoelectronic Functional Materials & Micro-Nano Devices, Department of Physics, Renmin University of China, Beijing 100872, PR China

[6] Department of Physics, Boston College, Chestnut Hill, MA, USA

[#]These authors contributed equally to this work.

[*]Correspondence to: hjgao@iphy.ac.cn; wangzi@bc.edu; wji@ruc.edu.cn; hlei@ruc.edu.cn

**In Weyl semimetals, charge density wave (CDW) order can spontaneously break the chiral symmetry[1,2], gap out the Weyl nodes[3], and drive the material into the axion insulating phase[4]. Investigations have however been limited since CDWs are rarely seen in Weyl semimetals[4,5]. Here, using scanning tunneling microscopy/spectroscopy (STM/S), we report the discovery of a novel unidirectional CDW order on the (001) surface of chiral crystal CoSi – a unique Weyl semimetal with unconventional chiral fermions[6-9]. The CDW is incommensurate with both lattice momentum and crystalline symmetry directions, and exhibits an intra unit cell π phase shift in the layer stacking direction. The tunneling spectrum shows a particle-hole asymmetric V-shaped energy gap around the Fermi level that modulates spatially with the CDW wave vector. Combined with first-principle calculations, we identify that the CDW is locked to the crystal chirality and is related by a mirror reflection between the two enantiomers of the chiral crystal. Our findings reveal a novel correlated topological quantum state in chiral CoSi crystals and raise the potential for realizing an axion insulator and exploring the unprecedented physical behaviors of unconventional chiral fermions.**

CDW is an electronic state in conducting materials where the charge density is periodically modulated. Experimentally, the formation of CDW is usually accompanied by increase in resistivity[4,10], gap opening around Fermi level[11-15], periodic modulations of the density of states, and distortions of the lattice due to electron-phonon coupling[11-13,16]. CDWs have been widely observed in crystals. In Weyl semimetals, the chiral Weyl fermions, with spin parallel to the momentum, are usually very stable, and CDW is rarely seen due to the low density of states at the Fermi level. There are several theoretical proposals for realizing CDWs in Weyl semimetals via chiral symmetry breaking[1,2] and electron-electron interactions[3,17]. Thus far, Weyl semimetals hosting CDWs are limited to very few quasi-one-dimensional materials[4,5].

CoSi is a diamagnetic[6,18], three-dimensional covalent compound in the chiral crystal family[19-25] described by a cubic B20 structure with space group of $P2_13$ (Fig. 1a). Its lattice structure has a well-defined handedness due to the lack of mirror and inversion symmetries[26], enabling two different growth enantiomers (referred to as CoSi and SiCo). The band structure of CoSi hosts two novel types of symmetry-enforced chiral Weyl nodes, the spin-1 and charge-2 nodes[6-8,27-30], which are widely separated in momentum space. Long Fermi arcs connecting their surface projections have been observed by angle-resolved photoemission spectroscopy (ARPES) measurements from different groups[7-9] and STM[31]. We report the discovery of a correlated Weyl fermion state – an incommensurate CDW whose wave vector is locked to the charity/handedness of CoSi and related by mirror reflection between the two distinct enantiomers.

A typical STM topography on our samples shows the atomically resolved square lattice of the CoSi (001) surface with pronounced periodic stripe-like modulations (Fig. 1b). The drift-corrected Fourier transformed (FT) image (see Methods) in Fig. 1c shows, in addition to the atomic Bragg peaks, two sharp peaks (pointed to by red arrows) at the unidirectional wave vector $Q_{cdw}$ corresponding to the stripe modulations. Zooming in to the first Brillouin zone (Fig. 1d), the magnitude of $Q_{cdw}$ is measured to be 0.32±0.01 π/a, while the angle between $Q_{cdw}$ and $k_x$ direction is 31±2°. As a result, the unidirectional charge modulations are incommensurate with both the lattice momentum and the high-symmetry lattice directions.

The tunneling conductance dI/dV (r, V) maps are subsequently measured. The dI/dV map in Fig. 2a acquired at V=-30 mV shows unidirectional stripe modulations. The corresponding Fourier transform (Fig. 2b) exhibits the same $Q_{cdw}$ as obtained from the topography. To confirm that the stripe modulations correspond to an electronic order such as the CDW, we took dI/dV maps at different energies (See Extended Data Fig. 1) and obtain the drift-corrected FT images from -30 to 30 mV. The normalized intensity of the FT along the blue dashed arrow in Fig. 2b is shown as a function of bias in Fig. 2c. Clearly, the wave vector of the stripe modulation is nondispersive and peaks around $Q_{cdw}$ at all energies. The higher harmonic of the modulation is also visible in Fig. 2c. This is in contrast to the energy dispersive quasiparticle interference patterns (QPI) due to the bulk or surface Fermi arc states (Extended Data Fig. 2), and suggests the stripe modulations originate from unidirectional charge order.

Further evidence for charge modulation can be seen by comparing the dI/dV maps between the occupied and unoccupied states. In Fig. 2d, the dI/dV maps taken at 24 mV (upper panel) and -24 mV (lower panel) show stripe modulations with a clear contrast reversal[11], as highlighted by the red and black dashed lines. This intensity reversal is consistent with the formation of a unidirectional Peierls CDW, where enhanced conductance for tunneling out of occupied states at a V < 0 in the region with charge accumulation implies enhanced conductance of a similar amplitude for tunneling into unoccupied state around –V over the region with charge depletion. Throughout the measurements, our data always exhibit such a contrast reversal between filled and empty states, which provide key evidence for the formation of a unidirectional CDW.

To probe the impact of the incipient CDW on the low-energy electronic states, the typical dI/dV spectra, obtained at the positions on (blue) and off (red) the stripes as marked in Fig. 1b, is plotted in Fig. 2e. The overall density of states (DOS) profiles are similar and show pronounced and broad peaks around -100 mV and 82 mV, suggesting high density of dispersive electronic states around these energies. The predicted chiral Fermi arc states on the (001) surface by density functional theory (DFT) span this energy range[6,32], as confirmed by ARPES[7-9], and are likely to contribute to these broad spectral peaks. The DOS decreases below these energies toward the Fermi level, in good agreement with the Weyl semimetallic nature of CoSi[6]. Remarkably, a particle-hole asymmetric energy gap emerges at low energies together with two sharp peaks at the gap edges and a V-shaped in-gap DOS having the minimum located near the Fermi level (Fig. 2e). Note that the energy gap is significantly larger than the spin-orbit coupling (SOC) induced band splitting[9,31], which has been observed to be less than 2 meV[33-35]. We thus speculate that the spectral gap around $E_F$ is the CDW gap due to the incommensurate charge modulation. Extracting the gap size from the peak to peak distance (Fig. 2e), we obtain $2\Delta_{cdw} \simeq 29$ meV. We note that the gap size ($2\Delta_{cdw}$) varies from 12 to 40 meV in different regions of three CoSi(001) samples.

Additional evidence supporting an incommensurate CDW gap in the DOS comes from the temperature dependence of the differential conductance. Fig. 2f displays the temperature evolution of the asymmetric dI/dV curves obtained at an on-stripe location, showing a gradual reduction of the CDW gap size with increasing temperature[36]. Our STM/S can clearly reveal stripe modulations up to 80 K (Extended Data Fig. 3). To gain insights into the melting temperature of the charge modulations, we performed low energy electron diffraction (LEED) measurements, which provide an estimate of the CDW transition temperature ~ 150 K (Extended Data Fig. 4).

The definitive evidence for the causal relations between the gap in the DOS spectrum and the unidirectional spatial modulations requires the spatial variations of the spectral gap. To this end, the dI/dV map at V=12 mV is shown in Fig. 2g. Extracting the gap value from the local tunneling spectrum at every point (Fig. 2g), we produce the gap map $2\Delta_{cdw}$(r) shown in Fig. 2h. Intriguingly, the gap sizes exhibit the same spatial modulations matching those of the striped modulations in the dI/dV map and the topography, confirming the unidirectional CDW nature of the spatial modulations and the low-energy V-shaped gap in the DOS.

We next demonstrate the unconventional properties of the unidirectional CDW, which manifest most remarkably across the step edges. We observe two types of terraces due to the absence of inversion symmetry[31]. The topmost atoms on different types of terraces can only overlap after a glide-mirror symmetry operation (Fig. 3a,b, Extended Data Fig. 5). The same type of terraces are connected by steps with heights of integer times the c-axis lattice constant *c* (even steps), while two terraces of different types are connected by step heights of half odd integer times *c* (odd steps). Such odd steps have been reported to play important roles in modifying the midgap step edge states in a topological crystalline insulator[37]. Hereafter we refer to the two types of terraces as terraces with different indices (***n*** or ***n+1/2***) (Fig. 3b). Fig. 3a shows a region with multiple terraces. The success step height between adjacent terraces is measured to be ~2.3 Å, about half of the *c*-axis lattice constant. Interestingly, the CDW stripes are in phase across even steps (Fig. 3c,d), while they are out of phase across odd steps (Fig. 3e,f). The in phase (out of phase) CDW pattern across even (odd) steps is reproducible in different CoSi samples. The near π phase shift across odd steps suggests strong inhomogeneous intra-unit cell charge distribution (Fig. 3e,f) with intricate correlations along the *c*-direction. This is consistent with the atomically resolved dI/dV linecut showing strong intra-unit cell conductance modulations (Extended Data Figs. 6,7).

Surprisingly, the dI/dV line shape associated with the CDW gap shows intriguing spatial evolution across an odd step (Fig. 3g-i). On the lower terrace in Fig. 3g, the asymmetric DOS has a larger spectral weight at the coherence peak on the unoccupied side at positive bias. This asymmetry reverses on the upper terrace, with higher spectral weight for coherence peak on the occupied side at negative bias. This reversal of line shape asymmetry is absent across even steps, where only modulations of the CDW gap size is observed (Extended Data Fig. 8). To the best of our knowledge, this behavior has not been observed in other CDW systems. In contrast to chiral crystal PdGa, another member of the space group $P2_13$ where different enantiomers have been found on two adjacent terraces across an odd step[21,38], we find no evidence of different enantiomers across odd steps in CoSi. Instead, the reversal of the asymmetry associated with the spectral weight transfer is closely related to the phase shift of CDW across odd steps, where the CDW order parameter picks up an additional π phase (Fig. 3e,f).

We next turn to the implication and the possible origin of the incommensurate unidirectional CDW in CoSi. Two recent transport studies have shown the existence of phonon-drag effect in CoSi[33,39], providing experimental evidence for significant electron-phonon interactions. The CDW can also be induced by electron-electron interactions when a large density of states is present due to partial Fermi surface nesting. In $(TaSe_4)_2I$, where similar CDW patterns have been recently observed[40], there are 24 pairs of Weyl points present around the Fermi energy. A linear combination of different nesting vectors has been argued to give rise to the experimentally observed short CDW wave vector[41]. The chiral crystal CoSi, on the other hand, has only one pair of well-separated Weyl nodes at Γ and R in the 3D Brillouin zone[6], and their spanning vector is far off in both magnitude and direction from the observed incommensurate CDW wave vector. As a result, if near nesting conditions are at play, the observed CDW is to originate from the individual small Fermi surfaces around the zone center and zone corners[6].

Further insights arise upon a closer examination of CDW direction with respect to the unique chiral crystal structure of CoSi. In each unit cell, the Co and Si atoms are distributed in four sublayers, labeled as 1 to 4 in Fig. 4a. The interlayer spacing between the top two sublayers is 80 pm, while the $2^{nd}$ and $3^{rd}$ sublayers are separated by a larger spacing of 131 pm. The observed stripes are oriented along path Co_1-Co_2-Si_2-Si_1-Co_1-Co_2…, as indicated by the red line in the top view of sublayers 1 and 2 in Fig. 4b, involving inter-sublayer tunneling. The other path highlighted by the grey line in Fig. 4b is inequivalent due to the absence of mirror and rotation symmetries associated with the chiral handedness of the crystal, when sublayers 3 and 4 are taken into account (Extended Data Fig. 9). This gives rise to the uniaxial behavior of the CDW, which is consistent with the absence of electronic stripe domains with the other orientation over structurally well-defined regions.

Occasionally, we observe a structural domain boundary on the surface, separating the observed area into two regions with different enantiomers (Fig. 4c). While the two regions share the same high-symmetry lattice directions, the orientations of the CDW wave vectors are remarkably different, as outlined by the blue and green dashed lines. The angle between those two CDW wave vectors in the two chiral structural enantiomers is ~69° (Extended Data Fig. 10), pointing to that the CDW is locked to the chiral handedness of the crystal enantiomers and the two regions can be transformed to each other by a mirror symmetry operation.

To develop more microscopic understanding of the chirality locking CDW wave vector in relation to the handedness of the CoSi crystal, we performed DFT calculations for the two different enantiomers (Fig. 4d,g). The corresponding electronic structures (Extended Data Fig. 11) show a flat band at the Γ point lying 15 meV above the Fermi level[9]. In real materials, this flat band has been observed by ARPES measurements to lie about 5 meV below the Fermi level[7,9], possibly due to electron correlation and/or internal electron doping effects. We therefore plot the two sets of constant energy contours at 15 meV above the calculated Fermi level in the 2D Brillouin zone, and regard them as the approximate "adjusted" Fermi surfaces for the two enantiomers (Fig. 4e,h). Each set of contours contains four lobes, and the two sets are related by mirror reflection. This is likely the reason behind the observed uniaxial electronic behavior. The joint density of states (JDOS) of the corresponding Fermi surface contours were calculated and plotted in Fig. 4f,i for both crystal enantiomers. Remarkably, the measured incommensurate CDW wave vectors, superimposed using the blue and green arrows, match unique sections of the JDOS with enhanced intensity for both enantiomers related by mirror symmetry. While more theoretical and experimental investigations are desirable to probe the microscopic origin, these results pointing to an electronic structure of CoSi harboring high density of quasiparticle scatterings at these wave vectors, making it conceivable that electronic correlations can drive the observed unconventional incommensurate CDW.

The combined STM experiments and DFT analysis suggest that the chirality locking CDW, with the sizable energy gap capable of reaching the Weyl node at Γ point below the Fermi level, has the potential to produce a mass gap for the unconventional chiral fermions. This raises the hope for realizing a highly nontrivial correlation-driven topological state governed by axion electrodynamics.

# Methods

**Single-crystal growth and surface treatment**

The quality of the sample surface has large influence on the clear observation of the CDW stripes. In order to obtain large and clean surface with atomically sharp steps, CoSi single crystal with good quality and sophisticated surface treatment is needed.

High-quality CoSi single crystals were synthesized by a chemical vapor transport method using $I_2$ as transport agent. High-purity Co powder (99.999%) and Si powder (99.99%) in 1:1 molar ratio were mixed with 150 mg $I_2$ and sealed in an evacuated quartz tube with a length of 110 mm and an inner diameter of 18 mm. The temperatures were set to 1000 °C (source side) and 900 °C (growth side), and kept for about 10 days, after which shiny single crystals (about 2 mm in size) with well-defined faces were obtained.

The crystal was then carefully grinded and polished along different high-symmetry orientations with different diamond lapping paper and polishing fluids by ALLIED™ MultiPrep polishing system. The desired crystal surface became shiny and showed no sign of scratches at 1000× optic magnification.

In order to minimize the chemical contamination on the surface, epoxy-free method was applied to mount the CoSi sample onto the STM sample holder. Sputtering and annealing procedure was carried out in an ultra-high vacuum chamber (base pressure better than $1 \times 10^{-9}$ torr), connected with STM chamber. After transferred into the vacuum chamber, the sample was repeatedly sputtered and annealed under 950 K. More than 50 cycles of sputtering and annealing had been applied for each sample to guarantee smooth and clean terraces with atomically sharp steps, which is crucial for the observation of the CDW order.

**STM/S experiments**

The STM/S measurements were operated at different temperatures in an ultra-low temperature STM system. Tungsten tips were etched chemically and calibrated on Au(111) surface before use. The *dI/dV* spectra and maps were obtained by a standard lock-in technique with a modulation voltage of 0.3 mV at 973.0 Hz. All the images and dI/dV maps were taken under 280 mK unless specifically mentioned. To eliminate the

STM tip-drift effect during long-time measurement in acquiring the topography and dI/dV maps, we apply the well-established Lawler-Fujita algorithm[42] and obtain a set of displacement fields and drift-corrected topography. The displacement fields are then applied to the dI/dV maps acquired simultaneously with the topography.

**DFT calculations**

Our DFT calculations were performed using the Perdew-Burke-Ernzerhof-generalized gradient approximation[43] (PBE) for the exchange-correlation potential, the projector augmented wave method[44] and a plane-wave basis set as implemented in the Vienna ab-initio simulation package[45] (VASP). A kinetic energy cutoff of 700 eV for the plane wave basis set was used for structural relaxation and electronic structures calculations. A k-mesh of 20×20×20 was adopted to sample the first Brillouin zone of the primitive unit cell of bulk CoSi. The shape and volume of each supercell were fully optimized and all atoms in the supercell were allowed to relax until the residual force per atom was less than $1\times10^{-3}$ eV·Å$^{-1}$. The theoretical lattice constant of the bulk cubic crystal was 4.43 Å, well consistent with the experimental values. Electronic bandstructures were calculated with considering SOC based on the two distinct enantiomers. The constant energy contours in the 2D Brillouin zone ($k_z = 0$) were plotted at 15 meV above the calculated Fermi level with an energy broadening of ±1 meV. The JDOS of corresponding constant energy contours were derived using Fourier transformations.

**Acknowledgements**

This work is supported by the Ministry of Science and Technology of China (2019YFA0308500, 2018YFA0305700 and 2018YFE0202600), the National Natural Science Foundation of China (61888102, 51991340, 52072401, 61761166009 and 11974422), the Chinese Academy of Sciences (XDB28000000, XDB07000000, XDB30000000, and YSBR-003), and Beijing Natural Science Foundation (Grant No. Z200005). Z.W. is supported by the US DOE, Basic Energy Sciences Grant No. DE-FG02-99ER45747. Calculations were performed at the Physics Lab of High-Performance Computing of Renmin University of China, Shanghai Supercomputer Center.

## Author contributions

H.-J.G. and G.L. designed the experiments; H.Y., Q.C., S.T. and H.L. synthesized the CoSi single crystals and polished the surfaces; G.L., P.J., G.H. and X.L. performed the STM/S experiment; G.L. performed the LEED experiment; C.W. and W. J. performed the DFT calculations; G.L., P.J., W.J., Z.W. and H.-J.G. analyzed the experimental data with inputs from all other authors; G.L., P.J. and G.H. plotted the figures. G.L., W.J., Z.W. and H.-J.G. wrote the manuscript with input from all other authors. H.-J.G. and Z.W. supervised the project.

## Competing interests

The authors declare no competing interests.

## Data availability

The data in this study are available from the corresponding authors on reasonable request.

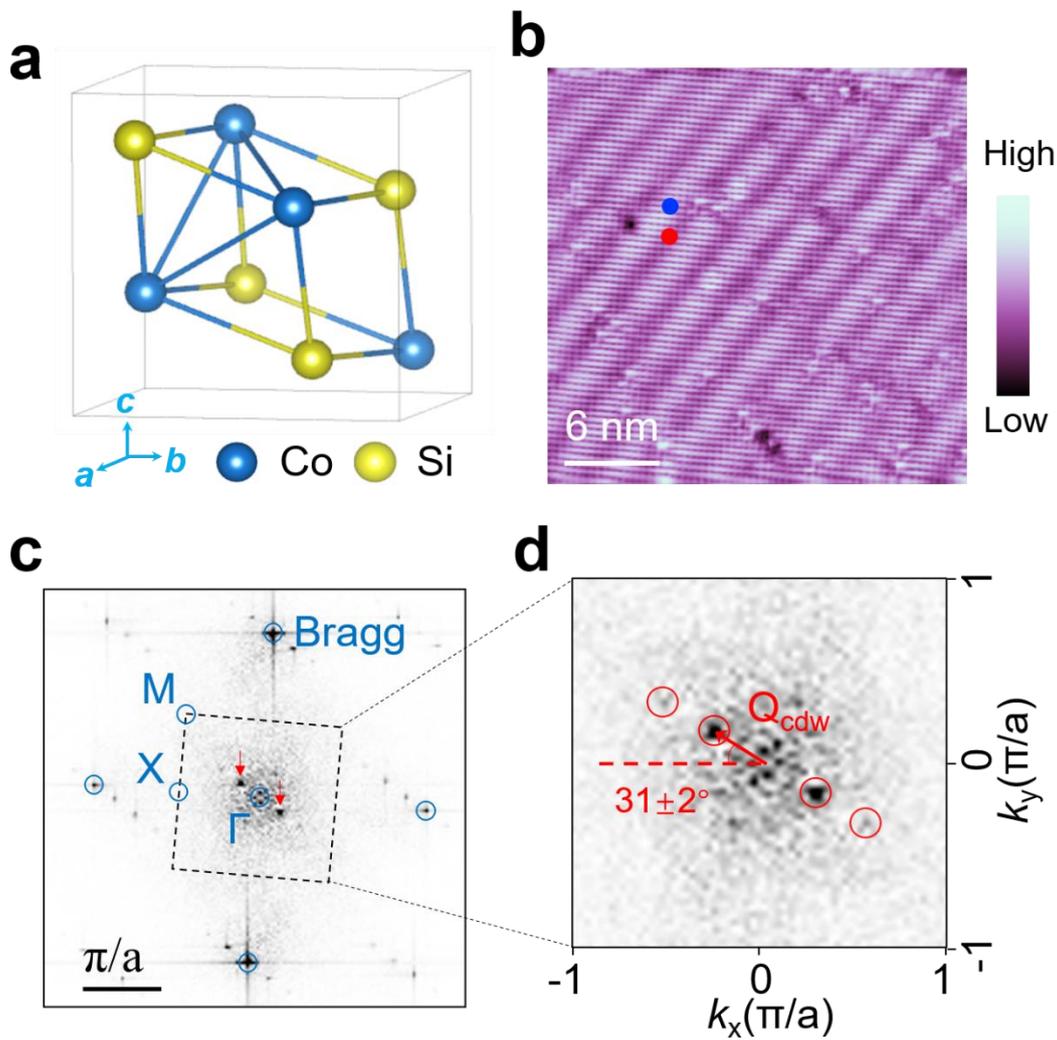

**Fig. 1 | Atomic structure, STM image and the FT images of the (001) surface of CoSi. a,** Crystalline structure of CoSi. **b,** Large-scale STM image of CoSi (001) surface showing stripe-like modulations. Scanning settings: $V_s$=-200 mV, $I_t$=0.4 nA. **c,** Fourier transformed image of **b**. The black dashed square outlines the 2D Brillouin zone of the surface. Bragg spots and high-symmetry points of the square lattice are outlined by blue circles. Around the Γ point, two sharp peaks associated with the stripe modulation can be differentiated, as highlighted by red arrows. **d,** Zoom-in image of the 2D Brillouin zone. The red arrow marks the CDW wave vector $Q_{cdw}$. The angle between $Q_{cdw}$ and $k_x$ direction is 31±2°.

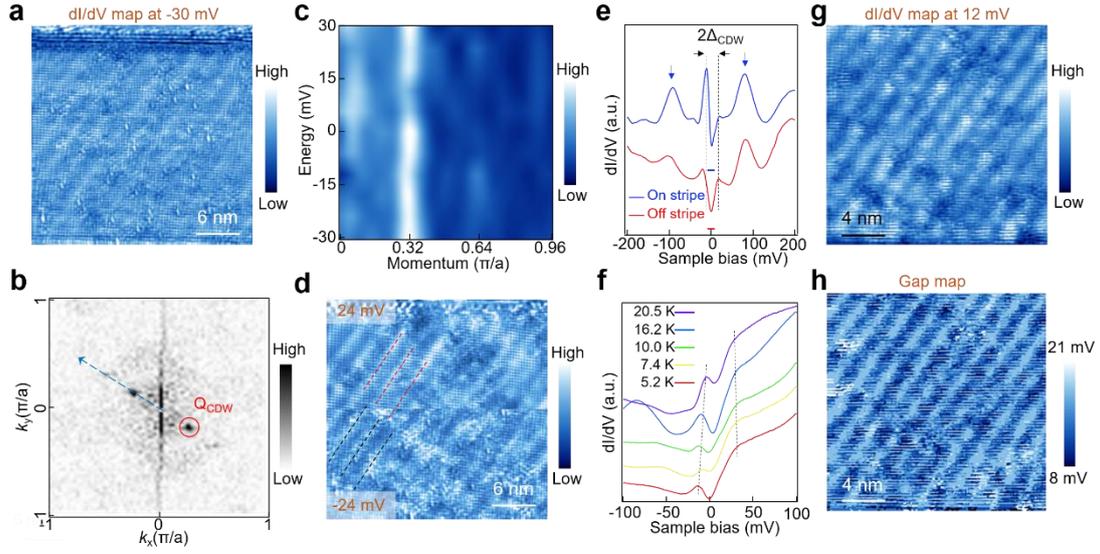

**Fig. 2 | Charge density wave on the (001) surface of CoSi. a,** dI/dV map of a surface region with the stripe modulations taken under -30 mV. **b,** 2D Brillouin zone of the surface withdrawn from the FT image of **a**. The CDW spot is outlined by a red circle. **c,** Normalized intensity profiles of the FT images along the blue dashed arrow in **b** as a function of bias of the region in **a**. The intensity peaks around $Q_{CDW}$ under different energies. The CDW wave vector is nondispersive with energy. **d,** dI/dV maps of the region in **a** taken at 24 mV (upper panel) and -24 mV (lower panel). The bright stripes of the upper (lower) are outlined by red (black) dashed lines. **e,** Typical dI/dV curves taken on (blue) and off (red) the bright stripes in Fig. 1**b**. Zero conductance for each spectrum is marked with a solid horizontal line. The blue arrows indicate two broad peaks in the spectra. The CDW gap $2\Delta_{CDW}$ is determined by the energy spacing between the left and right coherence peaks. **f,** Temperature dependence of the CDW gap. Each spectrum is taken at an on-stripe region in the same scanning area. The gap size gradually decreases with increasing temperature. **g,** dI/dV map at 12 mV taken at a region different from **a**. **h,** Gap map of the region in **g** acquired by recording the gap sizes at each pixel, showing a modulation of gap size by the CDW stripes.

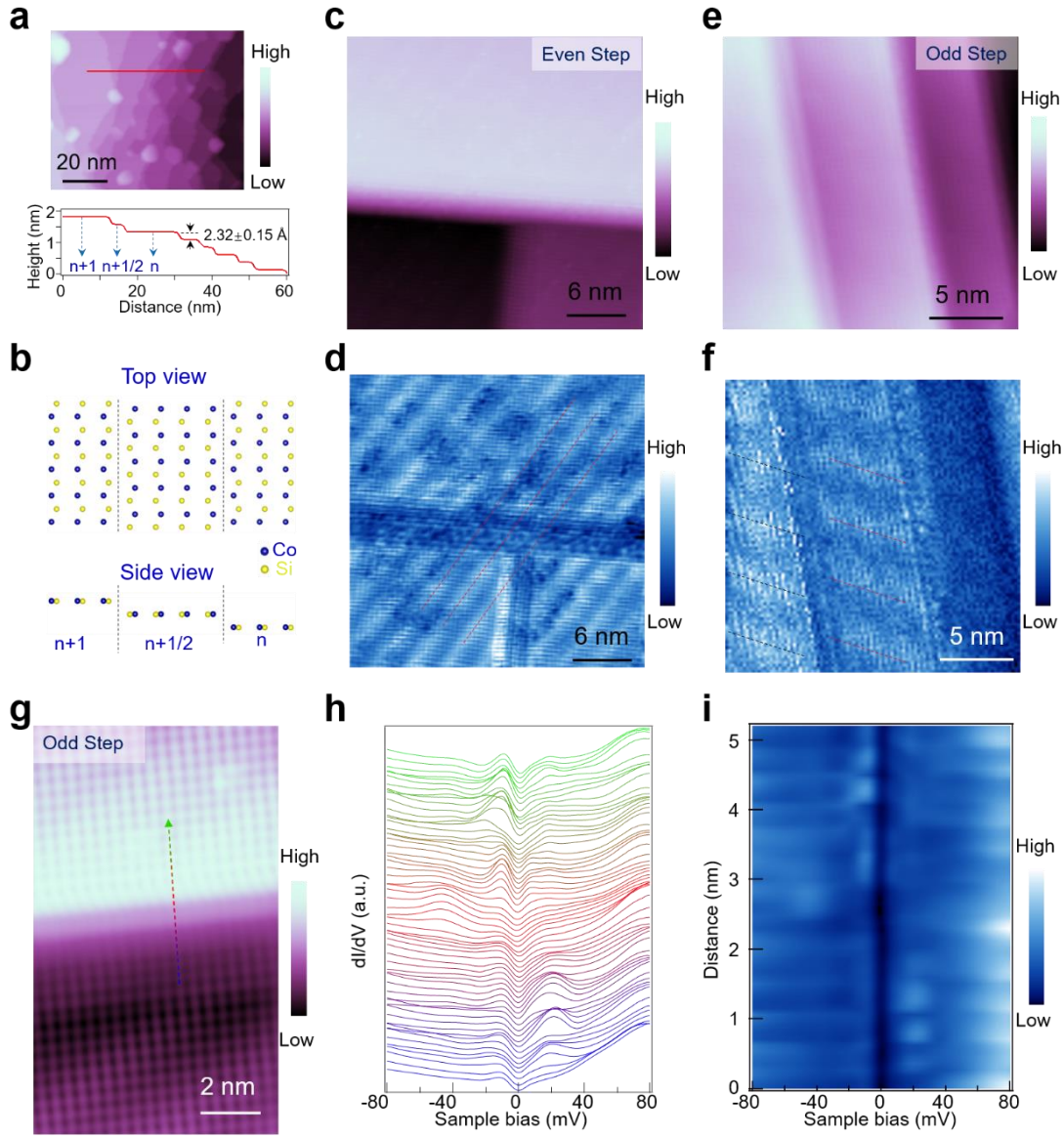

**Fig. 3 | Spatial evolution of CDW pattern and dI/dV spectra across steps. a,** Upper panel, a surface region showing terraces with different indices. Scanning settings: $V_s$=-1 V, $I_t$=0.6 nA. Lower panel, height profile along the red line in the upper panel, showing sequential steps with half of the lattice constant along *c* axis. **b,** Schematic drawing of the atomic arrangement of the terraces. The surface atoms on terraces possess n and n+1/2 indices, and are linked with glide-mirror symmetry. **c,d,** STM image (**c**) and dI/dV map under zero bias (**d**) of a region with steps. The height difference between the highest (upper part in **c**) and lowest (lower left part in **c**) terraces is 2 times of the c-axis lattice constant *c*. The CDW stripes are in phase across even steps, as outlined by red dashed lines. Scanning settings of **c**: $V_s$=-200 mV, $I_t$=0.1 nA. **e,f,** STM image (**e**) and dI/dV map under zero

bias (**f**) of a region with sequential odd steps with heights of 2.5$c$. The CDW stripes are out of phase across odd steps, as outlined by red and black dashed lines. Scanning settings of **e**: $V_s$=-200 mV, $I_t$=0.1 nA. **g,** STM image of an odd step. Scanning settings: $V_s$=500 mV, $I_t$=0.1 nA. **h,i,** Waterfall plot and intensity map of dI/dV spectra along the dashed arrow in **g**. On the lower terrace, the asymmetric CDW coherence peaks show larger spectral weight under positive bias, while under negative bias the spectral weight of the coherence peak gets stronger on the higher terrace. This asymmetry reverses across the odd step.

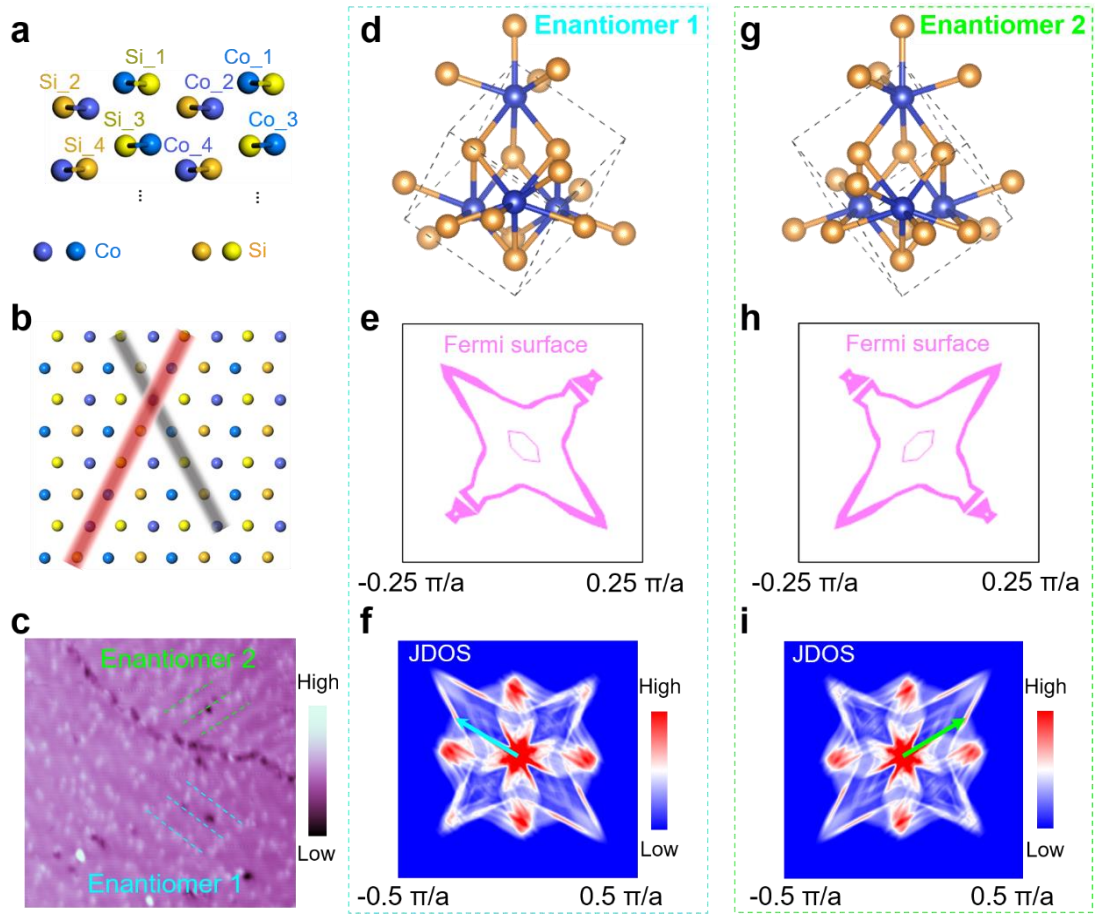

**Fig. 4 | Origin of chirality-locking, unidirectional CDW in CoSi. a,** Atomic structure of the four sub-layers in a unit cell of CoSi. **b,** Top view of the first and second sublayers. The red line outlines the electron hopping path of Co_1-Co_2-Si_2-Si_1-Co_1-Co_2…. The grey line outlines another path following the same atomic sequence in the two-sublayer model. These two paths are inequivalent with each other in bulk CoSi. **c,** Large-scale STM image showing coexistence of two different crystal enantiomers separated by a domain boundary. The blue and green dashed lines highlight the CDW stripes of enantiomers 1 and 2, respectively. The CDW orders are unidirectional in both enantiomers, but their CDW wave vectors point to different directions respectively. **d,g,** Atomic structures of crystal enantiomers 1 (**d**) and 2 (**g**). **e,h,** DFT revealed constant-energy contours of the Fermi surface at the adjusted Fermi level of enantiomers 1 (**e**) and 2 (**h**). These contours are only distributed around the Γ point in the *k*-space and the CDW was observed on the x-y plane of the crystal, thus, a slab at $k_z = 0$ with a range of ±0.25 π/a was plotted. **f,i,** Derived JDOS plots, with numerically enhanced contrast, of the DFT Fermi surface contours of enantiomers 1 (**f**) and 2 (**i**). The blue (**f**) and green

(**i**) arrows mark the experimentally observed CDW wave vectors in both direction and amplitude for enantiomers 1 and 2, respectively. The atomic structures, energy contours and JDOS of the two enantiomers show mirror symmetry with each other.